\documentclass[preprint,aps,showkeys]{revtex4-1}

\pdfoutput=1

\usepackage[english]{babel}
\usepackage[utf8]{inputenc}
\usepackage{amsmath}
\usepackage{graphicx}
\usepackage[space-before-unit,range-units = repeat]{siunitx}

\newcommand {\sqrbrc}[1]{\left[ #1 \right]}
\newcommand {\brc}[1]{\left( #1 \right)}
\newcommand {\txt}[1]{\text{#1}}
\newcommand {\avr}[1]{\left<{#1}\right>}
\newcommand {\abs}[1]{\left|{#1}\right|}
\newcommand {\ten}[1]{#1}

\begin{document}

\title{Generalized Navier Boundary Condition for a Volume Of Fluid approach
using a Finite-Volume method.}
\date{\today}
\author{A.M.P. Boelens}
\author{J.J. de Pablo}
\email{depablo@uchicago.edu}
\affiliation{University of Chicago}


\begin{abstract}
In this work, an analytical Volume Of Fluid (VOF) implementation of the Generalized
Navier Boundary Condition is presented based on the Brackbill surface tension
model. The model is validated by simulations of droplets on a smooth surface in
a planar geometry. Looking at the static behavior of the droplets, it is found
that there is a good match between the droplet shape resolved in the simulations and the
theoretically predicted shape for various values of the Young's angle.
Evaluating the spreading of a droplet on a completely wetting surface, the
Voinov-Tanner-Cox law ($\theta \propto \txt{Ca}^{1/3}$) can be observed. At
later times scaling follows $r \propto t^{1/2}$, suggesting spreading is limited
by inertia. These observations are made without any fitting parameters except
the slip length.
\end{abstract}
 
\maketitle

\section{Introduction}

For any multi-phase flow over a wall where dissipation at the contact line
becomes of the same order as the bulk dissipation, a good understanding of contact
line behavior is essential \citep{squires2005}. However, despite many
investigations \citep{dussan1979,degennes1985,deconinck2008,eral2013}, for applications
including adhesion of liquids to a solid surface \citep{furmidge1962}, transport
of liquid water in fuel cells \citep{zhu2008}, liquid infused surfaces
\citep{wexler2015}, and coating \citep{vandre2013}, contact line
behavior is still not well understood.
 
The physics of a static contact angle between a liquid and a gas on a
smooth solid surface is well established \citep{young1805,gauss1830}. However, real
surfaces are not completely smooth. They are not chemically homogeneous, and/or
have roughness. This causes static contact angle hysteresis and contact line
pinning, both of which are difficult to model. When looking at a moving contact
line instead of a static contact angle things get even more complicated. Dynamic
contact angle behavior is not well understood, even on a completely smooth solid
surface. The origin of this poor understanding of the moving contact line is
twofold: there is the contact line singularity problem, and the question
of how the contact angle depends on contact line velocity.

The contact line singularity problem was first identified by \citet{huh1971}.
While normally it is a good approximation to use the no-slip condition as
boundary condition on the wall, for corner flow this assumption causes the
viscous stress and pressure to scale as $r^{-1}$ and thus to diverge as $r \to
0$ at the contact line. Numerous methods have been proposed to resolve this
discontinuity or work around it. \citet{hocking1977} showed that, using domain
perturbation method in cylindrical coordinates, any slip-velocity model
\citep{navier1823,thompson1997} resolves the velocity singularity. Another
method to circumvent the contact line singularity has been to use the
Cahn-Hilliard-van der Waals model \citep{cahn1977} as the basis for either
diffusive interface models \citep{seppecher1996,jacqmin2000}, or for precursor
film models \citep{hervet1984,degennes1985}. Precursor film models use a
disjoining pressure \citep{derjaguin1955,teletzke1987,herring2010} to model the
van der Waals forces that cause the formation of a precursor film ahead of the
interface, removing the singularity. More exotic models have suggested local
shear-thinning \citep{cox1998} and non-constant surface tension
\citep{shikhmurzaev1993,shikhmurzaev1997} as possible solutions.

For the question of how the contact angle depends on the contact line velocity
there are also various models. Typically a distinction is made between the local
microscopic contact angle at the contact line and the macroscopic apparent
contact angle, which is observed in experiments. Due to experimental limitations
to access sufficiently small length scales, the apparent contact angle is
measured away from the contact line, and curvature of the interface causes this
angle to be different from the microscopic contact angle \citep{chen1989}.
Arguably, the easiest method to define the dynamic microscopic contact angle is
to simply assume it is fixed and the same as the static contact angle
\citep{eggers2004a}. Instead of a fixed dynamic contact angle, Molecular Kinetic
Theory (MKT) \citep{blake1969,blake2006} predicts a microscopic contact angle
which changes with the contact line velocity. The Voinov-Tanner-Cox
\citep{voinov1976,tanner1979,cox1986} law describes the relation between the
apparent contact angle and the microscopic contact angle. This law is based on
the assumption that it is possible to choose a length scale arbitrary close to
the contact line. This makes it impossible to identify a characteristic length scale
of the contact line geometry, and reduces its physics to a balance between
capillary and viscous stresses \citep{kafka1979}. Using the lubrication
approximation one can now derive the \citet{voinov1976} equation for some
specific asymptotic limits, and matching solutions of the Voinov equation with
the mesoscopic hydrodynamic solution further away from the wall then gives the
Cox-Tanner-Voinov law \citep{voinov1976,tanner1979,cox1986}.

Although there are a lot of different models to describe contact lines, there is
no consensus on what is the correct description of the physics. Many models have
multiple fitting parameters which can be tuned to give the same results
\citep{sibley2014}. Because impurities and surface heterogeneities have a large
effect on measurements, experiments are very difficult to reproduce. On top of
this, one needs access to microscopic length scales to get to the details and
the outcome of experiments on a macroscopic level only depends very weakly on
these small length scales \citep{bonn2009}. With the advent of Molecular
Dynamics (MD) \citep{koplik1988,koplik1989,thompson1989,thompson1990}
simulations, and new experimental techniques, such as Atomic Force Microscopy
(AFM) \citep{checco2006}, it is now possible to access length and time scales at
the contact line that were not accessible before. While MD simulations still can
only probe small systems for short times, a couple of fundamental discoveries
have been made. Although it had been argued that continuum models break down all together
at the contact line \citep{dussan1979}, it was found that both the Navier-Stokes
equations and Young's equation hold up even down to the nanometer scale
\citep{bocquet2010,seveno2013}. Furthermore, support was found for contact line slip
\citep{thompson1989}, precursor films \citep{he2003}, and non-static dynamic
microscopic contact angles \citep{shen1998}. Apart from simulations, precursor
films have also been found experimentally \citep{hardy1919,kavehpour2003}.

In addition to the experimental difficulties and often
contradicting findings, another reason that there is no consensus is that all
of the above mentioned models have some fundamental short comings and/or work in
different regimes. Further analysis of contact line slip models by
\citet{huh1977} showed that, even though stresses are not diverging anymore, the
pressure still shows a weak singularity i.e. the pressure diverges, but becomes
finite when integrated. While using a slip length model is sufficient from a
modeling perspective, a divergent pressure cannot be right from a physics
perspective. The precursor film on the other hand does
successfully work around the contact line singularity for both viscous stresses
and pressure. However, they are typically not seen under partial wetting
conditions \citep{snoeijer2013}. The different models for microscopic and
macroscopic contact angles have some limitations too. While a model which
describes the contact angle as a function of the capillary number might describe
dynamic contact angle hysteresis correctly, there is also static contact angle
hysteresis, which causes contact line pinning of non-moving contact lines. Any
model that lets the contact angle only depend on the local capillary number will
not be able to capture static contact angle hysteresis. Assuming a static angle
,on the other hand, does not capture static contact angle hysteresis, is not able
to properly describe the flow of a liquid over a chemically patterned surface
with different wetting properties, and does not accurately predict contact line
velocity \cite{yamamoto2013}.

In this work, a validation study of a Volume Of Fluid (VOF) implementation of a
contact line model called the Generalized Navier Boundary Condition (GNBC) is
presented. While the non-sharp interface of the VOF method implicitly
resolves the contact line singularity problem, even with a no-slip condition at
the wall \citep{seppecher1996,chen2000,sibley2013}, the question of what is the
right contact angle is still valid for this method. In addition to applying the
Navier slip condition at the wall, the GNBC uses the reduced Young's stress as a
restoring force when the contact angle deviates from its equilibrium value, and
is informed by Molecular Dynamics \citep{qian2003,gentner2004}. While previous
implementations used a friction factor to link the reduced Young's stress to a
contact line velocity \citep{gerbeau2009}, this approach does not work for the
VOF model, because of the implicit slip of the interface \citep{ashish2009}.
Instead, in this work, the reduced Young's stress is used directly as a source
term in the navier stokes equations \citep{mahady2015}. A consequence of using the GNBC is that the
contact angle no longer is a constraint imposed onto the system, but that the
contact angle is a self-selecting variable. Using this approach has a couple of
advantages over existing models. Because there is no enforcement of a model that
relates the contact angle to contact line speed, this model can reproduce static
contact angle hysteresis, without artificially fixing the position of the
contact line \citep{fang2008,dupont2010}. Additionally, a variable contact
angle model can describe flow over a chemically patterned surface
\citep{dupuis2004,wang2008}. While the Generalized Navier Boundary Condition has
been implemented for continuum simulations using a diffuse interface Cahn
Hillard method \citep{qian2006,ren2007,wang2008,cai2015}, using a Volume Of Fluid
approach has the advantage that many less grid points are needed to resolve the
interface \citep{ding2007}. In contrast to the body force term derived by
\citet{mahady2015}, the model presented in this work describes wetting of a dry
surface. Opposed to \citet{deganello2011}, the model presented here is a
analytical derivation of the line tension force.

In our work, we uncover the following findings: when looking at the static
behavior of droplets, it is found there is a good match between the droplet
shape resolved in the simulations and the theoretically predicted shape for
various values of the Young's angle. Investigating the spreading of a droplet on
a completely wetting surface, the Voinov-Tanner-Cox law
\citep{voinov1976,tanner1979,cox1986} ($\theta \propto \txt{Ca}^{1/3}$) can be
observed. Late time scaling follows $r \propto t^{1/2}$, suggesting spreading is
limited by inertia. These observations are made without any fitting parameters
except the slip length. 

\section{Theory}

\subsection{Body force reformulation of uncompensated Young's stress}

The traditional form of the Generalized Navier Boundary Condition (GNBC) used
in, for example, Arbitrary Lagrangian-Eulerian simulations \citep{gerbeau2009}, looks like:
\begin{equation}
  \beta \brc{\vec{u} - \vec{u}_{w}} \cdot \tau
+ \mu \brc{\nabla \vec{u} + \nabla \vec{u}^{T}} \hat{n}_{w} \cdot \tau
+ \sigma \brc{\cos{\theta} - \cos{\theta_{0}}} \hat{t}_{w} \cdot \tau \delta_{\txt{cl}}
=
  0
\label{eqn:gnbc}  
\end{equation}
Here the first two terms on the left hand side represent the Navier slip boundary
condition \citep{navier1823}, and the third term represents the unbalanced
Young's stress. $\beta$ is the slip coefficient, $\vec{u}$ is velocity,
$\vec{u}_{w}$ is the velocity of the wall, $\tau$ is any vector tangent to the
wall, $\mu$ is the dynamic viscosity, $\hat{n}_{w}$ is the normal of the wall
pointing outward, $\sigma$ is the surface tension coefficient, $\theta$ is the
dynamic contact angle, $\theta_{0}$ is the equilibrium Young's contact angle,
$\hat{t}_{w}$ is the vector tangent to the wall and normal to the contact line,
and the distribution $\delta_{\txt{cl}}$ is defined as \citep{gerbeau2009}:
\begin{equation}
  \avr{\delta_{\txt{cl}}, \psi}
=
  \int_{\txt{cl}} \psi d \lambda_{\txt{cl}}
\end{equation}
where $\psi$ is any smooth function, and $\lambda_{\txt{cl}}$ denotes the
Lebesgue measure (i.e. the length measure) on the contact line. The slip length
in equation \ref{eqn:gnbc} is equal to $l_{s} = \mu / \beta$.

Analogous to the methods of \citet{brackbill1992} a Volume Of Fluid expression is derived
for the above equation for the Generalized Navier Boundary Condition. However, instead of relating the
uncompensated Young's stress to a velocity using a slip coefficient, the
uncompensated Young's stress is modeled as an extra body force acting at the
contact line in the Navier-Stokes equation. The reason for this approach is that
the intrinsic contact line slip in a VOF code is large enough that converting
the contact line tension directly to a velocity on the wall does not move the
contact line. The Navier-slip boundary condition component of the above equation
is left unchanged in our implementation of the Generalized Navier Boundary
Condition. 

The goal of this derivation is to find an expression that rewrites the contact
line force as a body or volume force, so it can be treated as a momentum source term in
the Navier-Stokes equations. The first step is to find a relation, that rewrites
the line force $\vec{F}_{\tau L}$ as surface force $\vec{F}_{\tau A}$ acting on
the wall:
\begin{equation}
  \lim_{h \to 0} \int_{\Delta A_{w}} \vec{F}_{\tau A} \brc{\vec{x}} d^{2} \vec{x}
=
  \int_{\Delta L} \vec{F}_{\tau L} \brc{\vec{x}_{L}} d \vec{x}
\label{lineForce}
\end{equation}
The points, $\vec{x}_{L}$, form the contact line or triple point, $\Delta L$ is
the length of the line segment in a the small volume of integration $\Delta V$,
and $\Delta A_{w}$ is the side of $\Delta V$ that is part of the wall, and in
which the points $\vec{x}_{L}$ lay. An additional constraint for $\vec{F}_{\tau
A} \brc{x}$ is that it is zero outside of the interface region:
\begin{equation}
  \vec{F}_{\tau A} \brc{x}
=
  0
\txt{ for }
  \abs{
    \hat{n}_{2D} \brc{\vec{x}_{L}}
    \cdot
    \brc{\vec{x} - \vec{x}_{L}}
  }
\geq
  h
\end{equation}
where $\hat{n}_{2D} \brc{\vec{x}_{L}}$ is the normal to the contact line in the
plane of the wall, and $\delta \sqrbrc{\hat{n}_{2D} \brc{\vec{x}_{L}} \cdot
\brc{\vec{x} - \vec{x}_{L}}}$ describes the plane $\Delta A_{w}$.

Consider a system of two fluids, fluid $1$, and fluid $2$, separated by an
interface, and define a discontinuous function, $c \brc{\vec{x}}$, to distinguish
between the two phases:
\begin{equation}
  c \brc{\vec{x}}
=
  \left\{
    \begin{array}{ll}
      c_{1}                           & \txt{in liquid 1}  \\
      \avr{c} = \brc{c_{1} + c_{2}}/2 & \txt{on interface} \\
      c_{2}                           & \txt{in liquid 2}
    \end{array}
  \right.
\end{equation}
An example of such a function would be the density of two different
incompressible liquids, $\rho_{1}$, and $\rho_{2}$. In this case the position of
the contact line can be found with:
\begin{equation}
  \rho \brc{\vec{x}_{L}}
=
  \avr{\rho}
\end{equation}
In order to change this problem from a boundary value problem on the contact
line to an approximate continuous model, one can define a continuous function
$\tilde{c}$, which varies smoothly over thickness $h$ going from $c_{1}$ to
$c_{2}$ , $c_{1} \le \tilde{c} \le c_{2}$. $h$ is a length scale of the order of
the grid size $\Delta x$, and defines the width of the transition region from
$c_{1}$ to $c_{2}$. The two functions $c$ and $\tilde{c}$ are related through
the interpolation function $\mathcal{P} \brc{\vec{x}}$:
\begin{equation}
  \tilde{c} \brc{\vec{x}}
=
  \frac{1}{h^{2}} \int_{A} c \brc{\vec{x}'} \mathcal{P} \brc{\vec{x}'-\vec{x}} d^{2} \vec{x}
\end{equation}
which is normalized as:
\begin{equation}
  \int_{A} \mathcal{P} \brc{\vec{x}} d^{2} \vec{x}
=
  h^{2}
\end{equation}
is bounded as:
\begin{equation}
  \mathcal{P} \brc{\vec{x}}
=
  0 
  \txt{ for }
  \abs{\vec{x}}
\ge
  h/2
\end{equation}
is differentiable, and decreases monotonically with $\abs{\vec{x}}$. The continuous
function is defined such that:
\begin{equation}
  \lim_{h \to 0} \tilde{c} \brc{\vec{x}}
= 
  c \brc{\vec{x}}
\end{equation}
i.e. the function $\tilde{c} \brc{\vec{x}}$ approaches $c \brc{\vec{x}}$ as the
interface thickness goes to zero. $\tilde{c}$ is differentiable because $\mathcal{P}$ is, and:
\begin{equation}
  \nabla_{2D} \tilde{c} \brc{\vec{x}}
=
  \frac{1}{h^{2}} \int_{A} c \brc{\vec{x}'} \nabla_{2D} \mathcal{P} \brc{\vec{x}'-\vec{x}} d^{2} \vec{x}
\end{equation}
where $\nabla_{2D}$ is the two-dimensional gradient in the plane of the wall.
Using Gauss' theorem and the realization that $c \brc{\vec{x}}$ is constant
within each fluid, the above integral can be written as:
\begin{equation}
  \nabla_{2D} \tilde{c} \brc{\vec{x}}
=
  \frac{\sqrbrc{c}}{h^{2}} \int_{L} \hat{n}_{2D} \brc{\vec{x}_{L}} \mathcal{P} \brc{\vec{x}-\vec{x}_{L}} d L
\label{gauss}
\end{equation}
where $\sqrbrc{c} = c_{2} - c_{1}$, thus converting the surface integral to a
line integral. To pull the normal out of the integral, its weighted mean is
calculated. Since $\mathcal{P}$ is bounded, its maximum contribution to the line
integral is $\mathcal{O} \brc{h}$. Integral \ref{gauss} can thus be approximated
as:
\begin{equation}
  \frac{1}{h^{2}} \int_{L} \hat{n}_{2D} \brc{\vec{x}_{L}} \mathcal{P} \brc{\vec{x}-\vec{x}_{L}} d L
\approx
  \frac{1}{h^{2}} \hat{n}_{2D} \brc{\vec{x}_{L0}} \int_{L} \mathcal{P} \brc{\vec{x}-\vec{x}_{L}} d L
+
  \mathcal{O} \brc{\frac{h}{R}}
\label{approximate}
\end{equation}
where $\vec{x}_{L0}$ is the point on $L$ closest to $x$, and $R$ is the radius
of the contact line. 

The integral in equation \ref{approximate} can be bounded by:
\begin{equation}
  \frac{1}{h} \int_{L} \mathcal{P} \brc{\vec{x}-\vec{x}_{L}} d L
\le
  \mathcal{P} \brc{\vec{x}-\vec{x}_{L0}}
\end{equation}
where in the limit $h \to 0$, $\mathcal{P} \brc{\vec{x}-\vec{x}_{L0}}$ is zero
everywhere except for $\vec{x} = \vec{x}_{L0}$. Taking the corresponding limit
of $\nabla_{2D} \tilde{c} \brc{\vec{x}}$ over the interface gives:
\begin{equation}
  \lim_{h \to 0} \int \hat{n}_{2D} \brc{\vec{x}_{L0}} \cdot \nabla_{2D} \tilde{c} \brc{\vec{x}} dx
= 
  \sqrbrc{c}
\end{equation}
As $h$ goes to $0$, $\nabla_{2D} \tilde{c} \brc{\vec{x}}$ is thus equivalent to:
\begin{equation}
  \lim_{h \to 0} \nabla_{2D} \tilde{c} \brc{\vec{x}}
=
  \hat{n}_{2D} \sqrbrc{c} \delta_{2D} \sqrbrc{\hat{n}_{2D} \cdot \brc{\vec{x} - \vec{x}_{L}}} 
=
  \nabla_{2D} c \brc{\vec{x}}
\label{delta}
\end{equation}

Because the Brackbill surface tension already takes care of the $\cos(\theta)$
component of the reduced Young's stress, only the $\cos{\theta_{0}}$ component
needs to be modeled. Using the delta function, the contact line force can now be
written as a surface force as follows: 
\begin{align}
  \int_{L} \vec{F}_{\tau L} \brc{\vec{x}_{L}} d L 
&= 
  \int_{A} \vec{F}_{\tau L} \brc{\vec{x}} \delta_{2D} \sqrbrc{\hat{n}_{2D} \brc{\vec{x}_{L}} \cdot \brc{\vec{x} - \vec{x}_{L}}} d^{2} \vec{x} \nonumber \\
&= 
  \int_{A} \sigma \cos{\theta_{0}} \hat{n}_{2D} \brc{\vec{x}} \delta_{2D} \sqrbrc{\hat{n}_{2D} \brc{\vec{x}_{L}} \cdot \brc{\vec{x} - \vec{x}_{L}}} d^{2} \vec{x}
\end{align}
converting the line integral over the contact line into an integral over the
wall surface. Equation \ref{delta} can be used as an approximation for the delta
function when the interface has a finite thickness. Substitution gives:
\begin{equation}
  \int_{\Delta L} \vec{F}_{\tau L} \brc{\vec{x}_{L}} d L 
=
  \lim_{h \to 0} \int_{\Delta A_{w}} \sigma \cos{\theta_{0}} \frac{\nabla_{2D} \tilde{c} \brc{\vec{x}}}{\sqrbrc{c}} d^{2} \vec{x}
\label{surfaceForce} 
\end{equation}
and comparing equation \ref{lineForce} with \ref{surfaceForce}, the surface force
$\vec{F}_{\tau A}$ can be identified as:
\begin{equation}
  \vec{F}_{\tau A}
=
  \sigma \cos{\theta_{0}} \frac{\nabla_{2D} \tilde{c} \brc{\vec{x}}}{\sqrbrc{c}}
\end{equation}
As a last step, the surface integral needs to be converted to a volume integral.
Since $\nabla_{2D} \tilde{c}$ is independent of the distance away from the wall,
integrating over $\Delta V$ is the same as multiplying $\vec{F}_{\tau A}$ with
mesh size, i.e. $\Delta A_{w} / \Delta V$:
\begin{equation}
  \vec{F}_{\tau V}
=
  \vec{F}_{\tau A} \frac{\Delta A_{w}}{\Delta V}
=
  \sigma \cos{\theta_{0}} \frac{\nabla_{2D} \tilde{c}
  \brc{\vec{x}}}{\sqrbrc{c}} \frac{\Delta A_{w}}{\Delta V}
\label{eqn:volumeForce}
\end{equation}
where $\Delta A_{w}$ is the surface area of the wall in volume $\Delta V$.

While the above equation has the correct limiting behavior, it was found that,
due to the diffuse nature of the interface, at too low resolution the interface gets
pulled apart. To counter this phenomenon, a modification is proposed to localize
the contact line force more to the interface:
\begin{equation}
  g \brc{\tilde{c}}
=
  H \brc{\pi/2-\theta_{0}} \frac{5}{4} H \brc{\tilde{c}-\frac{1}{5}}
+ H \brc{\theta_{0}-\pi/2} \frac{5}{4} H \brc{\frac{4}{5}-\tilde{c}}
\end{equation}
Where $H(x)$ is the Heaviside step function. Depending on the value of the
Young's angle, this function truncates the contact line force. For $\theta_{0} <
\pi/2$ this means that the contact line does not get pulled apart, while for
$\theta_{0} > \pi/2$ this prevents the gas phase from being pulled into the
droplet at the contact line. The value $5/4$ is to normalize the function. The
new function for the contact line restoring force at low resolution thus becomes:
\begin{equation}
  \vec{F}_{\tau V}
=
  \vec{F}_{\tau A} \frac{\Delta A_{w}}{\Delta V} g \brc{\tilde{c}}
\label{eqn:volumeForceCor}
\end{equation}
where the same definition is used for $\vec{F}_{\tau A}$ as in equation
\ref{eqn:volumeForce}.

\subsection{Numerical implementation}
This section focuses on the numerical implementation of the above derived
equation in the Volume Of Fluid (VOF) solver that comes with OpenFOAM
\citep{jasak1996,openfoam}. This involves implementing the uncompensated
Young's stress and the Navier slip boundary condition. In this code the general
phase parameter $\tilde{c} \brc{\vec{x}}$ is called $\alpha$, and has the
following properties: 
\begin{equation}
  \alpha
=
  \left\{
  \begin{array}{ll}
  0      & \txt{in phase 1}    \\
  (0, 1) & \txt{on interface}    \\
  1      & \txt{in phase 2}
  \end{array}
  \right.
\end{equation}
Phase parameter $\alpha$ is stored as a separate field, just like velocity and
pressure, and its evolution is calculated using the following transport equation:
\begin{equation}
  \frac{\partial \alpha}{\partial t}
+ \nabla \cdot \brc{\alpha \vec{v}}
+ \nabla \cdot \brc{\alpha \brc{1 - \alpha} \vec{v}_{lg}}
=
  0
\end{equation}
where $\vec{v} = \alpha \vec{v}_{l} + \brc{1 - \alpha} \vec{v}_{g}$ is the phase
averaged velocity, and $\vec{v}_{lg} = \vec{v}_{l} - \vec{v}_{g}$ is the
velocity difference between the liquid and the gas phase. This equation is
equivalent to a material derivative, but rewritten to minimize numerical
diffusion \cite{rusche2002}.

The volume fraction is used to calculate phase-averaged densities, velocities,
and viscosities, which are used in the momentum balance
\begin{equation}
  \frac{\partial \rho \vec{v}}{\partial t}
+ \nabla \cdot \brc{\rho \vec{v} \otimes \vec{v}}
=
- \nabla p
+ \nabla \cdot \brc{\mu \nabla \vec{v}}
+ \rho \vec{g}
+ \vec{f}_{\txt{st}}
- \vec{f}_{\txt{cl}}
\end{equation}
and the continuity equation:
\begin{equation}
  \nabla \cdot \vec{v}
=
  0
\end{equation}
In the above equations $\rho$ is the density, $\vec{v}$ is the velocity, $t$ time, $p$ is
pressure $\mu$ is the viscosity, $g$ is gravity, $\vec{f}_{\txt{st}}$ is the
surface tension force, $\vec{f}_{\txt{cl}}$ is the contact line tension force,  and
$\otimes$ is the dyadic product. The density $\rho$, velocity $\vec{v}$, and
viscosity $\mu$, are all phase averaged using $\alpha$.

The surface tension force is calculated using the expression:
\begin{equation}
  \vec{f}_{\txt{st}}
=
  \sigma_{\txt{st}} \kappa \nabla{\alpha}
\end{equation}
where $\sigma_{\txt{st}}$ is the surface tension coefficient,
\begin{equation}
  \kappa
=
- \brc{\nabla \cdot \vec{n}}
\end{equation}
is the curvature of the interface, and
\begin{equation}
  \vec{n}
=
  \frac{\nabla \alpha}{\abs{\nabla \alpha}}
\end{equation}
is the normal of the interface \cite{brackbill1992}.

Using $\alpha$, the line tension force defined in equation
\ref{eqn:volumeForceCor} is rewritten as:
\begin{equation}
  \vec{f}_{\txt{cl}}
=
  \sigma_{\txt{st}} \cos{\theta_{0}} \;
  g \brc{\alpha}
  \frac{\Delta A_{w}}{\Delta V}
\end{equation}
and $\nabla_{2D} \alpha$ is implemented as:
\begin{equation}
  \nabla_{2D} \alpha 
=
  \nabla \alpha - \brc{\hat{n}_{w} \cdot \nabla \alpha} \hat{n}_{w}
\end{equation}
where $\hat{n}_{w}$ again is the normal of the wall pointing outward. The
equilibrium Young's angle $\theta_{0}$ can be defined uniquely for any grid cell along
the wall, so an arbitrary wettability pattern can be created. $\Delta A_{w}$ and
$\Delta V$ are properties of the mesh that can be accessed directly in OpenFOAM.
The contact line tension source term is solved explicitly along with the surface
tension source term.

The Navier slip condition is implemented as: 
\begin{equation}
  \vec{v}
=
  l_{s} \brc{\ten{I} - \hat{n}_{w}^{2}} \nabla \vec{v}
\end{equation}
where $l_{s}$ is the slip length $\ten{I}$ is the identity matrix, and
$\hat{n}_{w}^{2}$ is the dyadic product of the wall normal, $\hat{n}_{w}$, with
itself. Using this formulation, the velocity perpendicular to the wall is always
set to zero.

\section{Results}

The first section of the results focuses on the static behavior of the
implementation of the Generalized Navier Boundary Condition presented above. The
second section covers the dynamic contact angle behavior. To speed up simulation
times all results are for 2D droplets in a planar geometry, and using the
symmetry of the system only half of the droplet is simulated. All simulations are
performed without gravity and represent water droplets in air at room
temperature.

\subsection{Static}

\begin{figure}
\includegraphics[width=0.6\textwidth]{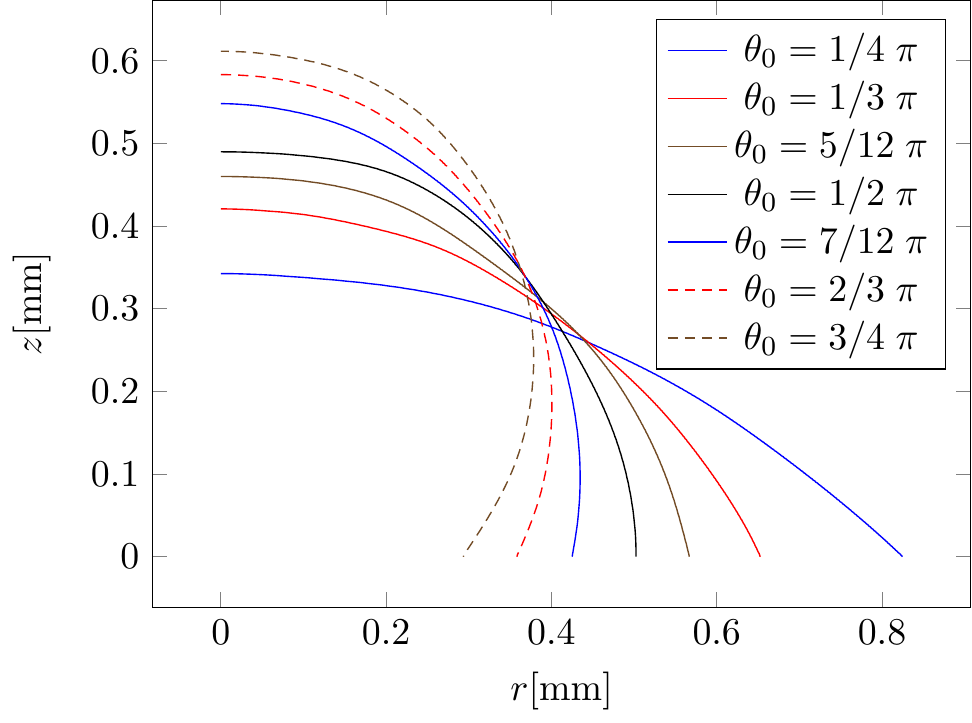}
\caption{The interface of various droplets at time $0.1 \si{\s}$ \label{figure:contourEquil}}
\end{figure}
Figure \ref{figure:contourEquil} shows the interface ($\alpha = 0.5$) of various droplets
with different Young's angles at a resolution of $512 \times 256$ in a box of
$1.5 \si{\milli\m} \times 0.75 \si{\milli\m}$. The half
droplets in the figure have a surface area of about $0.2 \si{\milli\m^{2}}$ (i.e. a radius of
$0.5 \si{\milli\m}$ when the Young's angle is $\pi/2$). As initial condition
these droplet where given their equilibrium shape, and it can be seen that after
a simulation time of $0.1 \si{\s}$ the droplets have maintained their shape.

\begin{figure}
\includegraphics[width=0.6\textwidth]{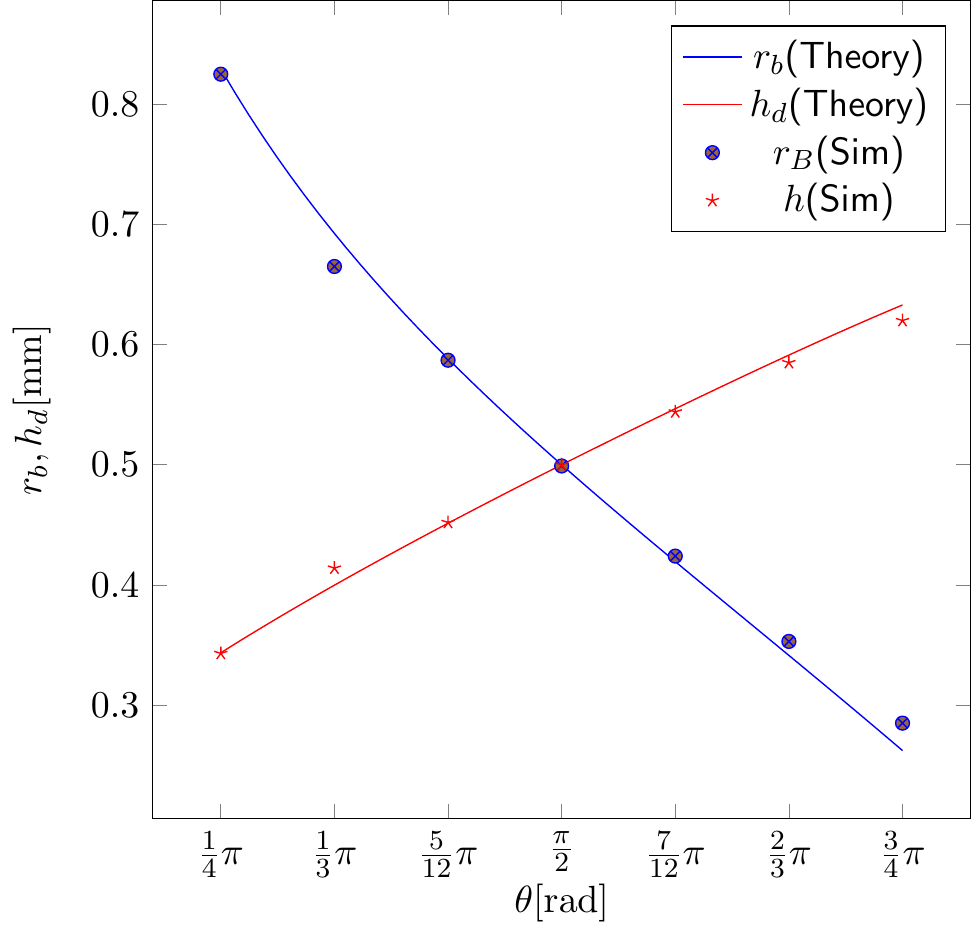}
\caption{Radius of the base of the droplet and droplet height as a function of
different contact angles.\label{figure:angleRadiusEquil}}
\end{figure}
To further quantify these droplets, figure \ref{figure:angleRadiusEquil} shows the
height and radius of the base of the droplets as a function of the Young's angle.
The radius of the base of the droplet $r_{B}$ is calculated as:
\begin{equation}
r_{b} =  R \sin{\theta_{0}},
\end{equation}
and the height of the droplet is equal to:
\begin{equation}
h_{d} = R \brc{1 - \cos{\theta_{0}}}.
\end{equation}
In the above equations:
\begin{equation}
R = \sqrt{\frac{2 A}{\theta_{0} - \cos{\theta_{0}} \sin{\theta_{0}}}}
\end{equation}
is the radius of the droplet, $\theta_{0}$ is the Young's angle, and $A$ is the
surface area of the droplet. While there are small deviations between the
theoretically predicted droplet shapes and the simulations, both match well.

\begin{figure}
\includegraphics[width=0.6\textwidth]{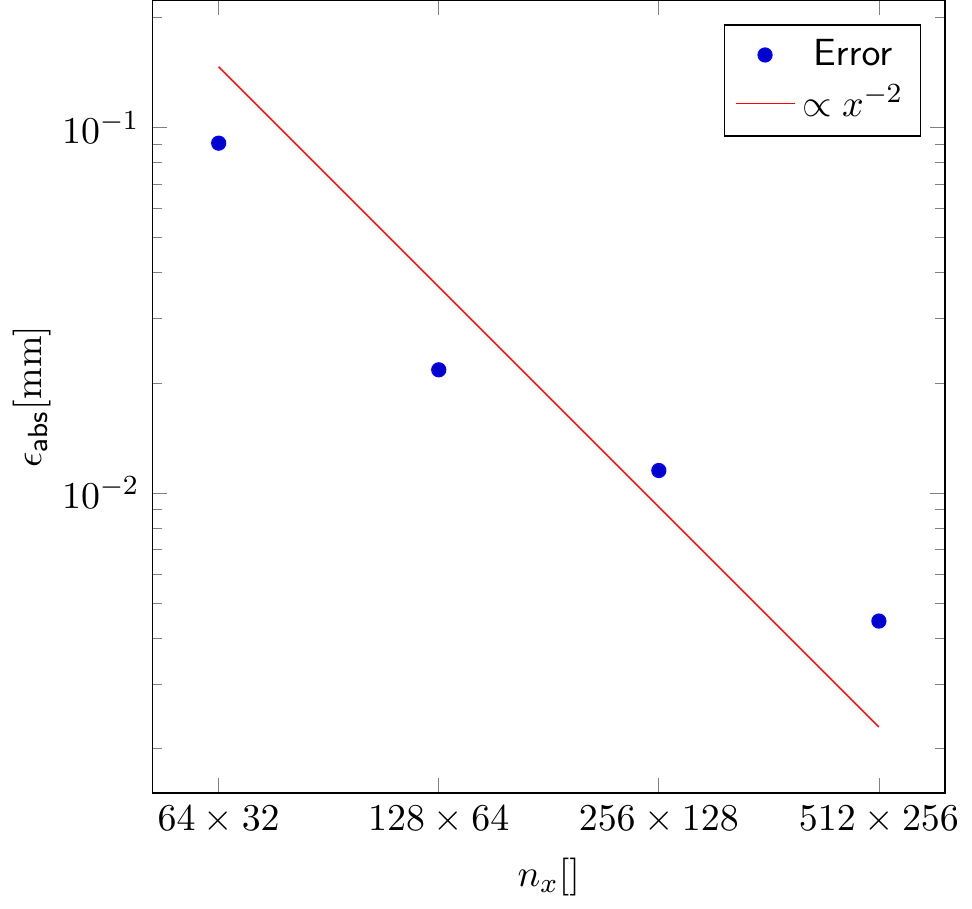}
\caption{Error in the radius of the base of the droplet as a function of
resolution for a contact angle of $\theta_{0} = \pi/4$.
\label{figure:resolution}}
\end{figure}
The convergence of the error as a function of resolution is shown in figure
\ref{figure:resolution}. The x axis shows the resolution. The values of the time steps at these resolutions are:
$0.1 \si{\mu \s}$, $0.05 \si{\mu \s}$, $0.025 \si{\mu \s}$, and $0.01 \si{\mu
\s}$, to keep the Courant number of the same order between simulations. The y
axis shows the absolute error, $\epsilon_{\txt{abs}}$, between the
theoretical value for the radius of the base of the droplet and the simulated
value for a Young's angle of $\theta_{0} = \pi/4$. The graph shows that the code
is close to 2nd order accurate.

\begin{figure}
\includegraphics[width=0.6\textwidth]{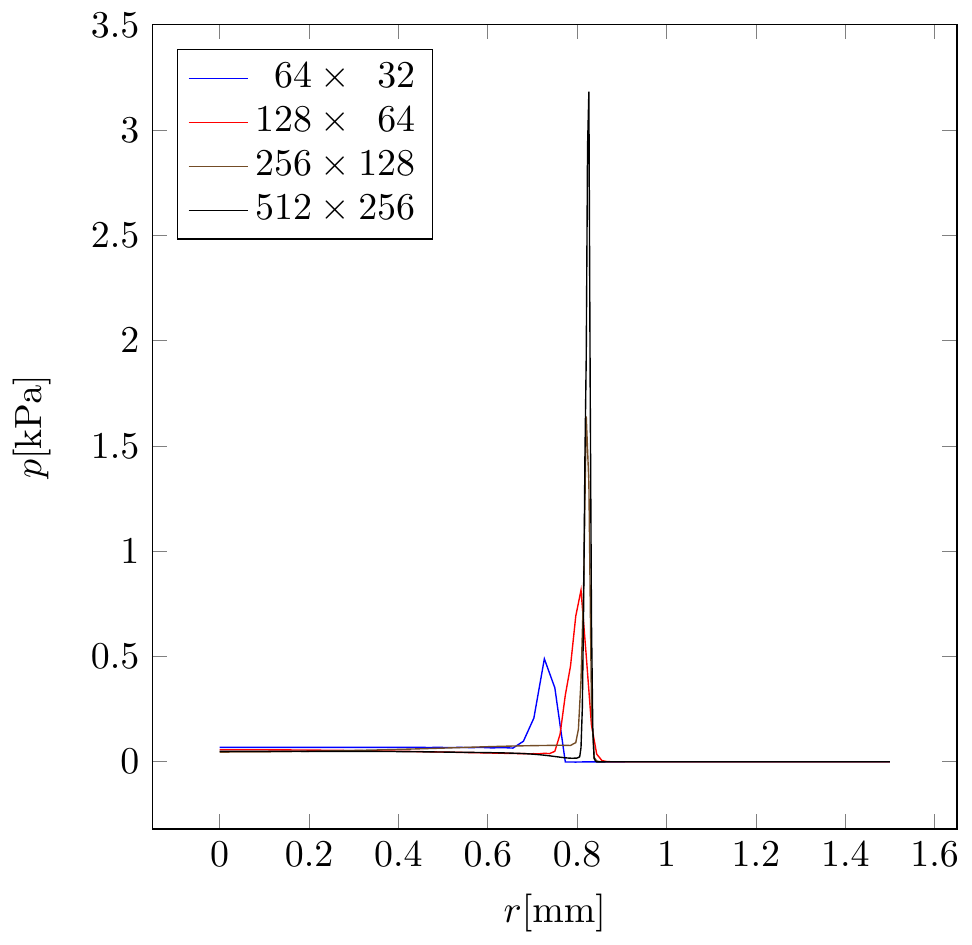}
\caption{\label{figure:pressureEquil} Pressure at the wall at $t = 50 \si{\milli
\s}$ for different resolutions and  a contact angle of $\theta_{0} = \pi/4$}
\end{figure}
Because the interface gets thinner as resolution increases the Volume Of Fluid
method is mesh dependent. As can be seen in figure \ref{figure:pressureEquil},
this results in the pressure peak on the surface getting sharper with increasing
resolution. The increased pressure inside the drop, left of the pressure peak,
is the Laplace pressure.

\begin{figure}
\includegraphics[width=0.6\textwidth]{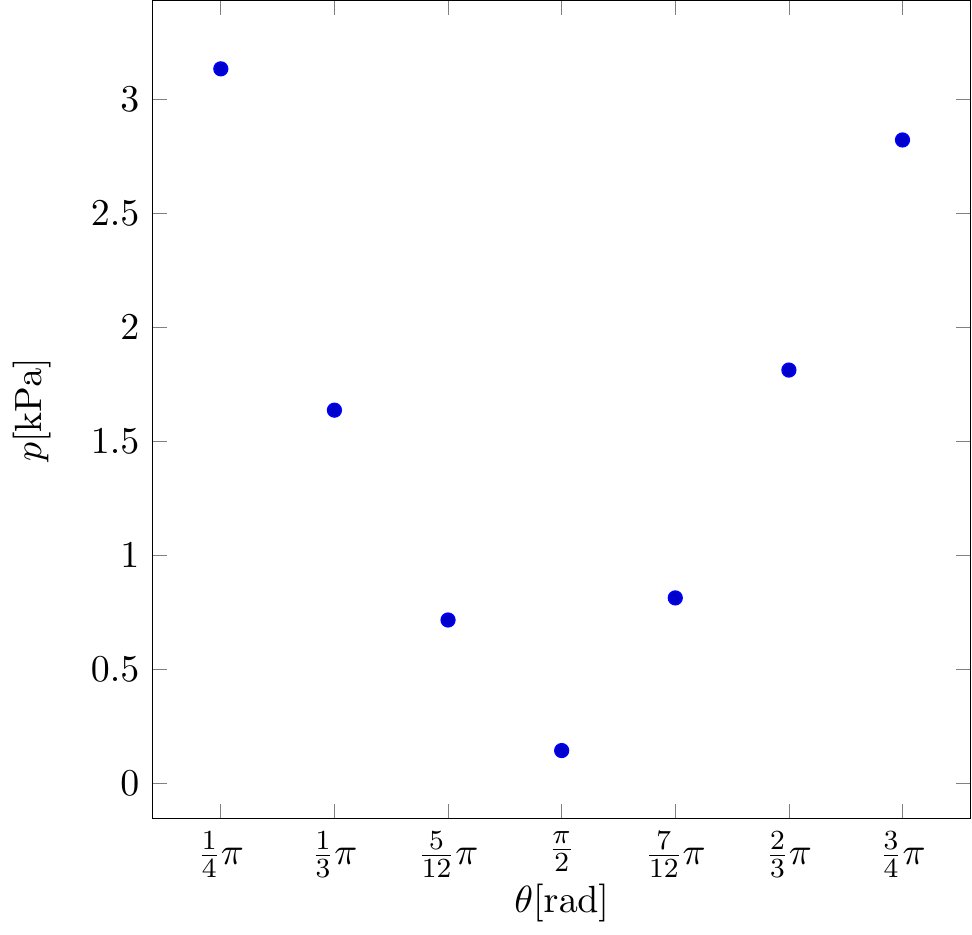}
\caption{Maximum of pressure peak at the contact line as a function of different
contact angles.\label{figure:anglePresEquil}}
\end{figure}
The reason there is a pressure peak at the contact line for a stationary droplet
in the first place can be seen in figure \ref{figure:anglePresEquil}. This
figure shows the value of the pressure peak as a function of the Young's angle.
From the figure it is clear that there is a strong dependence of the pressure
peak on the Young's angle. Since the line tension force only acts parallel to
the surface, this suggest that the pressure peak is the result of the surface
tension force calculated by the \citet{brackbill1992} model. If the
vertical component of the reduced Young's stress also were implemented, an extra
term, proportional to  $\sigma \sin{\theta_{0}}$, would have been present in the
plot as an additional contribution to the pressure peak. Since we are only
concerned about solid surfaces, this term was not incorporated in the model.

\begin{figure}
\includegraphics[width=0.6\textwidth]{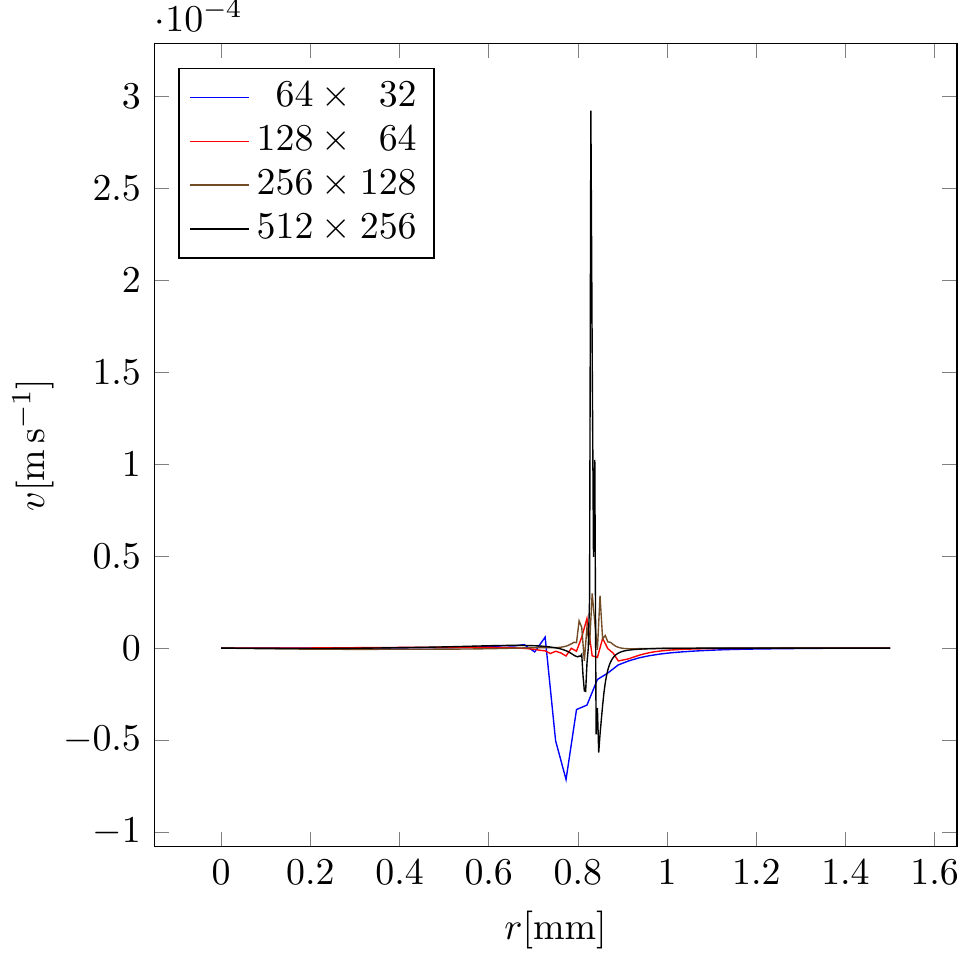}
\caption{\label{figure:velocityEquil} Velocity at the wall at $t = 50 \si{\milli \s}$ for different resolutions}
\end{figure}
Due to the large pressure peak there is also a small residual slip velocity at
the contact line, which can be observed in figure \ref{figure:velocityEquil}. As
was the case with the pressure, this velocity peak gets sharper at higher
resolutions. As can be seen in figure \ref{figure:angleRadiusEquil} the residual
slip velocity does not negatively affect the shape of the drop. However,
spurious currents are a well known issue of Volume Of Fluid
solvers \cite{deshpande2012} and, if needed, can be controlled by making the
time step sufficiently small. 

\begin{figure}
\includegraphics[width=0.6\textwidth]{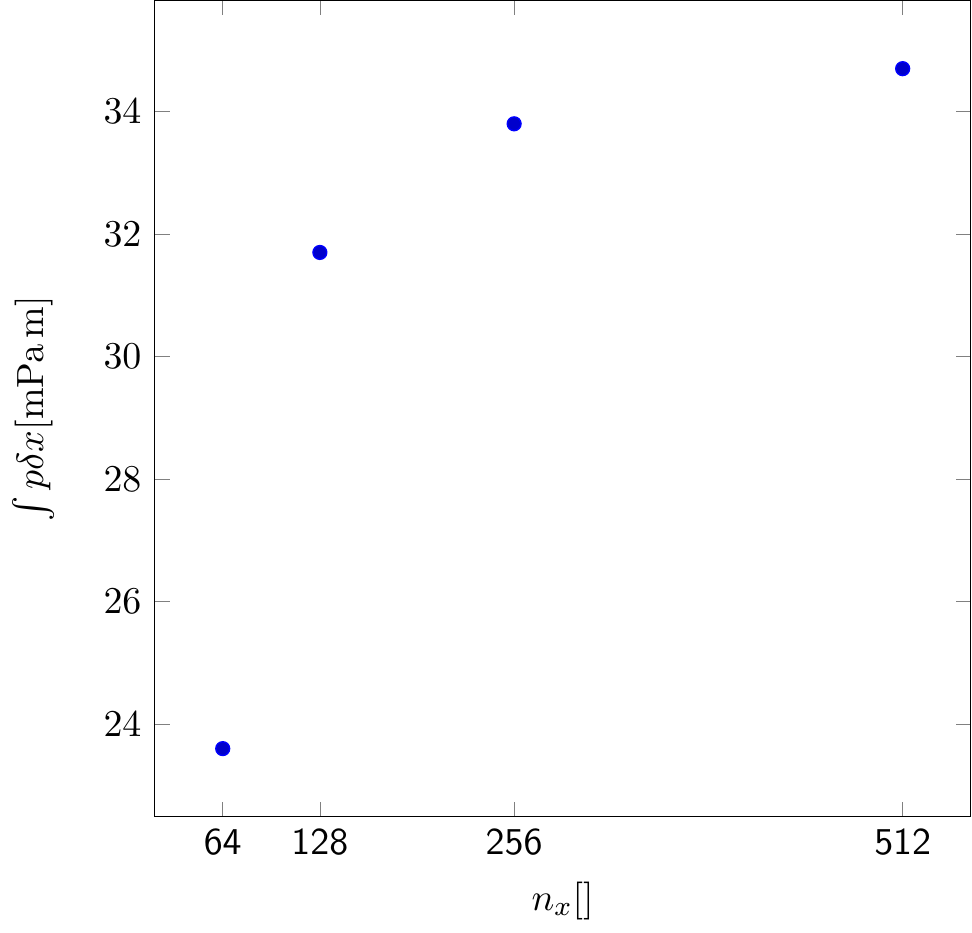}
\caption{Integral of pressure peak at the contact line as a function of
resolution.\label{figure:resolutionPressure}}
\end{figure}
\begin{figure}
\includegraphics[width=0.6\textwidth]{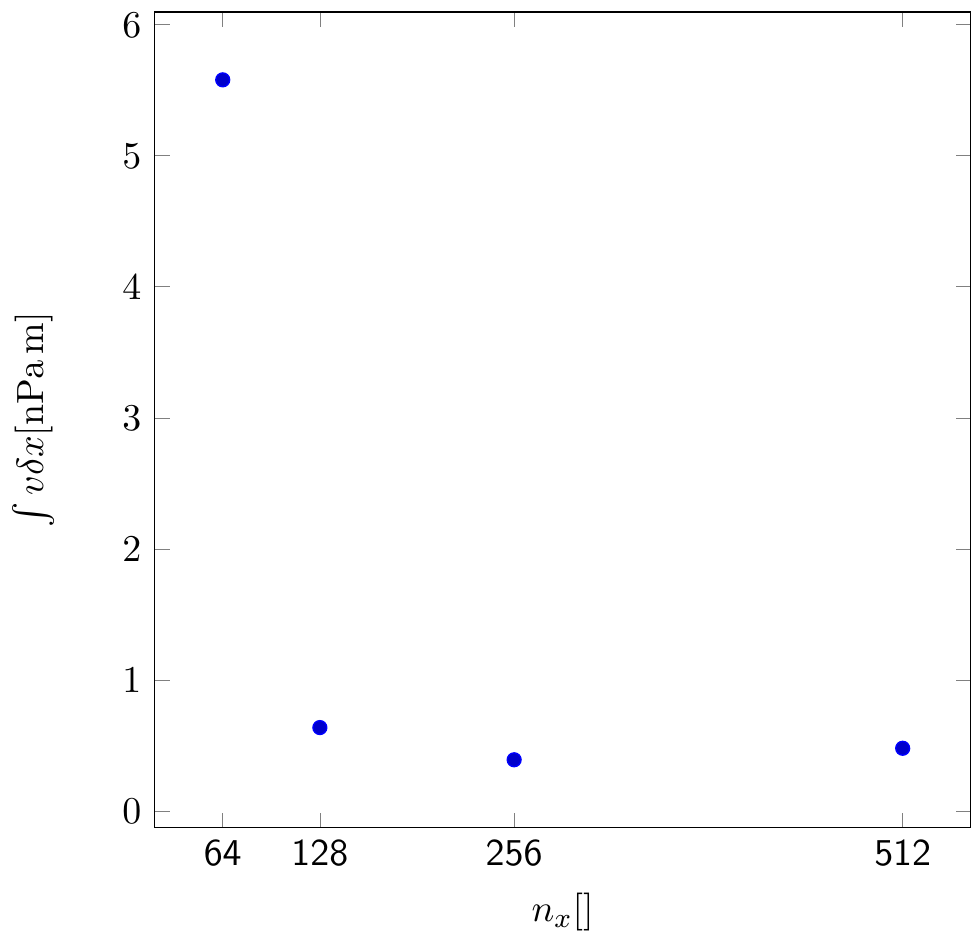}
\caption{Integral of slip velocity at the contact line as a function of
resolution.\label{figure:resolutionVelocity}}
\end{figure}
To investigate to what extent the pressure and velocity peaks affect the
solution they are both integrated over the wall for various resolutions. Figure
\ref{figure:resolutionPressure} shows this integral for the pressure, and figure
\ref{figure:resolutionVelocity} for the slip velocity. The integration limits in
both plots are from $\alpha = 0.01$ to $\alpha = 0.99$. For the integral over
the pressure peak this approach makes sure that the integral is not affected by
the Laplace pressure inside the droplet, and the velocity integral uses the same
limits to be consistent with the pressure. It can be appreciated how the
pressure integral converges to a constant value and the velocity integral
approaches zero as resolution increases, showing a convergent solution for both
the pressure and velocity.

\subsection{Dynamic}

For the validation of the dynamic case, the starting point is again a 2D droplet
in planar geometry. However, in this case, the equilibrium Young's contact angle
is set to $\theta_{0} = 0$, and the spreading behavior as a function of
time is investigated. 

\begin{figure}
\includegraphics[width=0.6\textwidth]{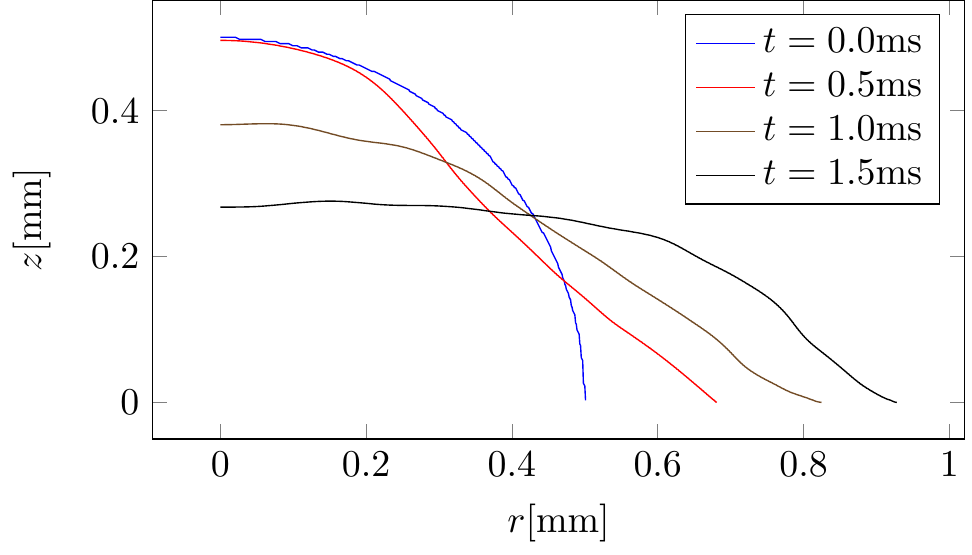}
\caption{\label{figure:contourTanner} Interface of a droplet at different times
at a resolution of $512 \times 256$.}
\end{figure}
Figure \ref{figure:contourTanner} shows the interface
($\alpha = 0.5$) at various times for a box of $1.5 \si{\milli\m} \times 0.75 \si{\milli\m}$
and with a resolution of $512 \times 256$. As expected the
droplet keeps spreading until the edge of the simulation box is reached and the
simulation is stopped.

\begin{figure}
\includegraphics[width=0.6\textwidth]{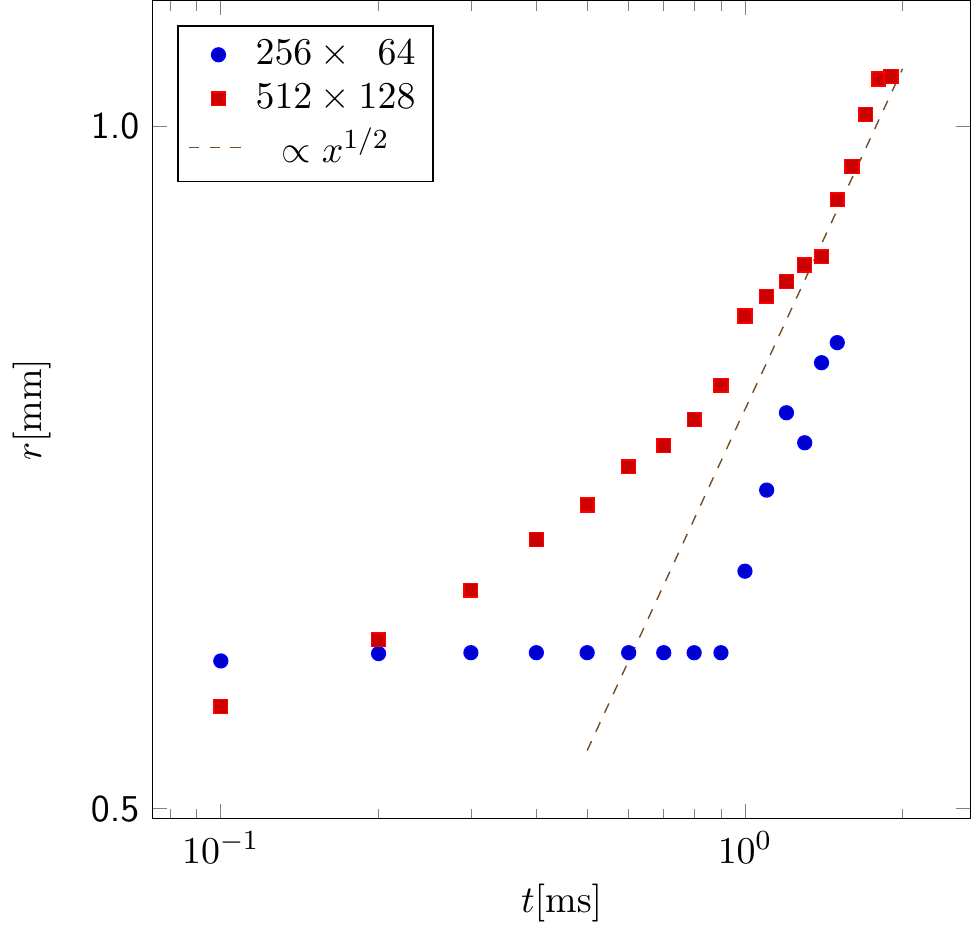}
\caption{\label{figure:diameterTanner} Radius of a droplet as a function of time.}
\end{figure}
Figure \ref{figure:diameterTanner} shows how the droplet spreads by showing the
radius of the droplet as a function of time for two different resolutions of $256
\times 128$ and $512 \times 256$. For lower resolutions the contact line was
pulled apart and the results are not shown in the graph. For reference, the power
law $r \propto t^{1/2}$ is also shown. The $r \propto t^{1/2}$ power law
describes the late time spreading behavior for low viscosity axisymmetric droplets \citep{biance2004,ding2007b}. At a resolution of $256 \times 128$ initially the contact
line hardly moves, but then the spreading radius as a function of time becomes
proportional to: $r \propto t^{1/2}$. At the larger resolution of $512 \times
256$ it can be seen that the scaling also converges to $r \propto t^{1/2}$.

\begin{figure}
\includegraphics[width=0.6\textwidth]{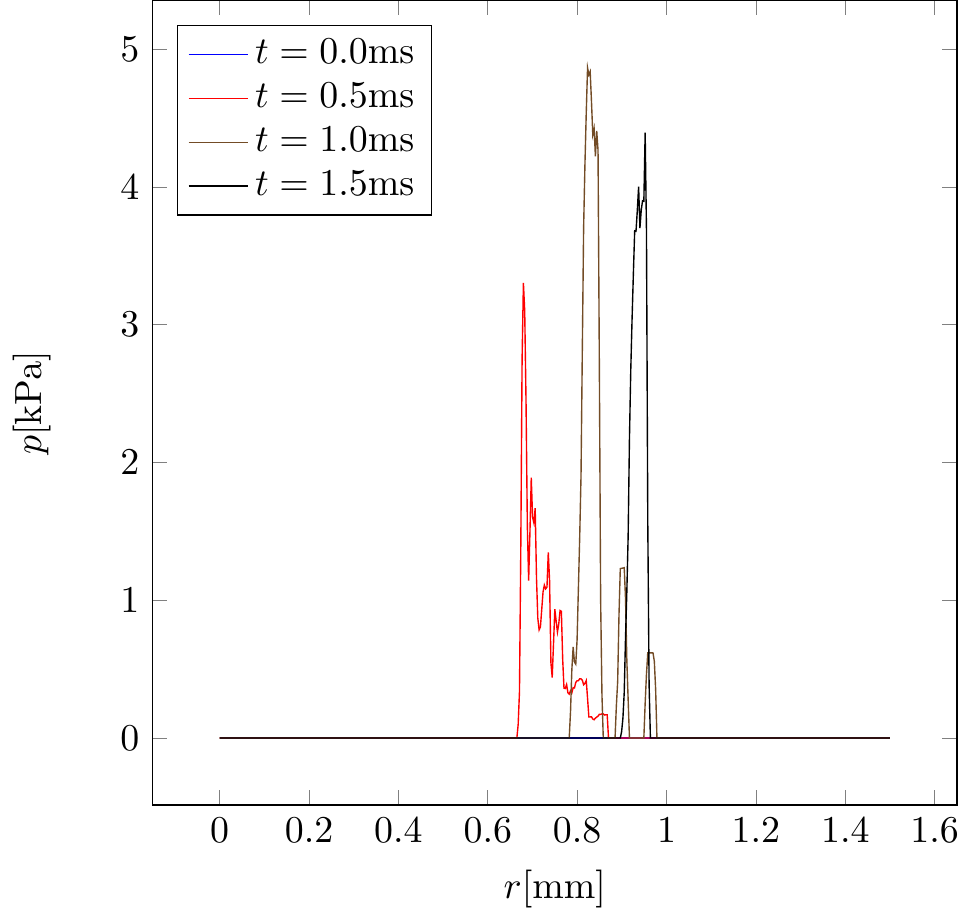}
\caption{\label{figure:pressureTanner} Pressure at the wall for various times at
a resolution of $512 \times 256$}.
\end{figure}
\begin{figure}
\includegraphics[width=0.6\textwidth]{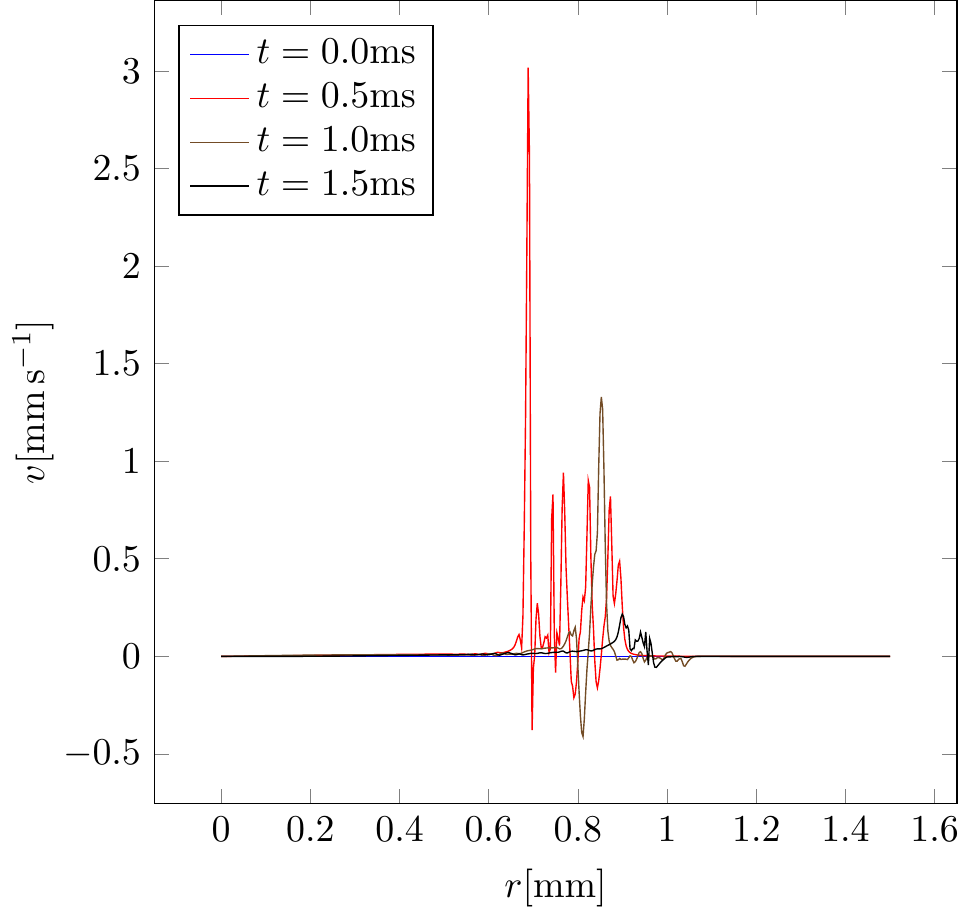}
\caption{\label{figure:velocityTanner} Velocity at the wall for various times at
a resolution of $512 \times 256$}.
\end{figure}
Figures \ref{figure:pressureTanner} and \ref{figure:velocityTanner} show the
pressure profile and velocity at the wall for various times. The
pressure shows a sharp peak already, but especially at $t=0.5 \si{\milli\s}$
there is quite some noise in both the curves for pressure and velocity. The
peaks are expected to become even sharper with increasing resolution
\citep{afkhami2009}. While it is known that for the Navier-slip boundary
condition pressure is divergent \citep{devauchelle2007}, whether the reduced
Young's stress provides a cut-off mechanism for the pressure has not yet been
analytically determined.

Because of the difference in the spreading curves in figure
\ref{figure:diameterTanner} an additional simulation was performed at a much
larger resolution. The simulation domain for this simulation is
$0.4\si{\milli\m} \times 0.2\si{\milli\m}$ with a resolution of $256 \times 128$,
but the mesh is refined at the wall to better capture the curvature at the
contact line. The grid cells at the wall are about $10 \si{\nano\m}$ cubed. The
droplet has an initial radius of $0.15\si{\milli\m}$ and makes an angle to
$\theta = \pi/4$ with the surface. The function $g$ which was used in the above
simulations to keep the contact line tension localized was omitted in this
simulation because it is not needed at larger resolutions.

\begin{figure}
\includegraphics[width=0.6\textwidth]{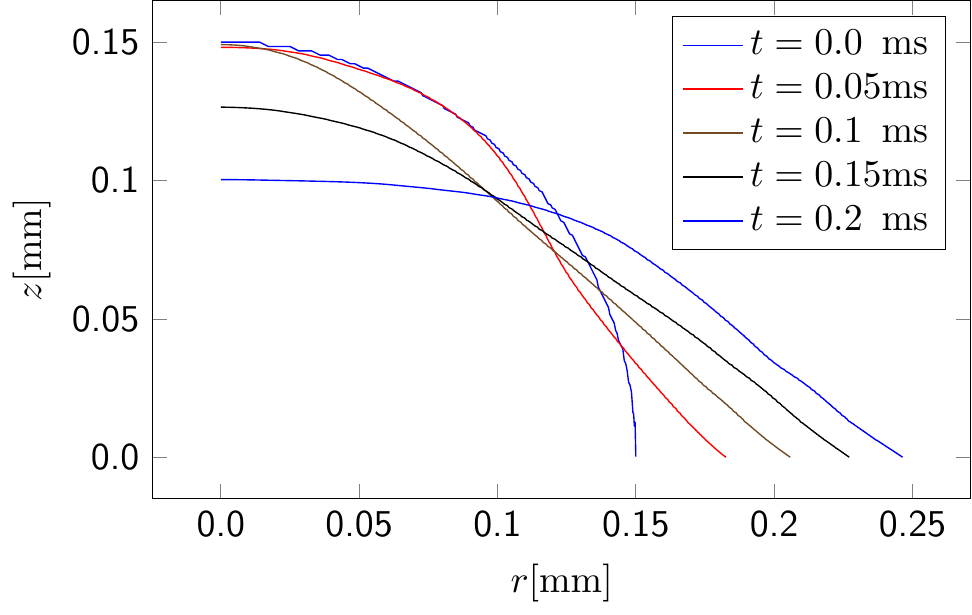}
\caption{\label{figure:contourTannerRes} Interface of a droplet at different
times at high resolution.}
\end{figure}
\begin{figure}
\includegraphics[width=0.6\textwidth]{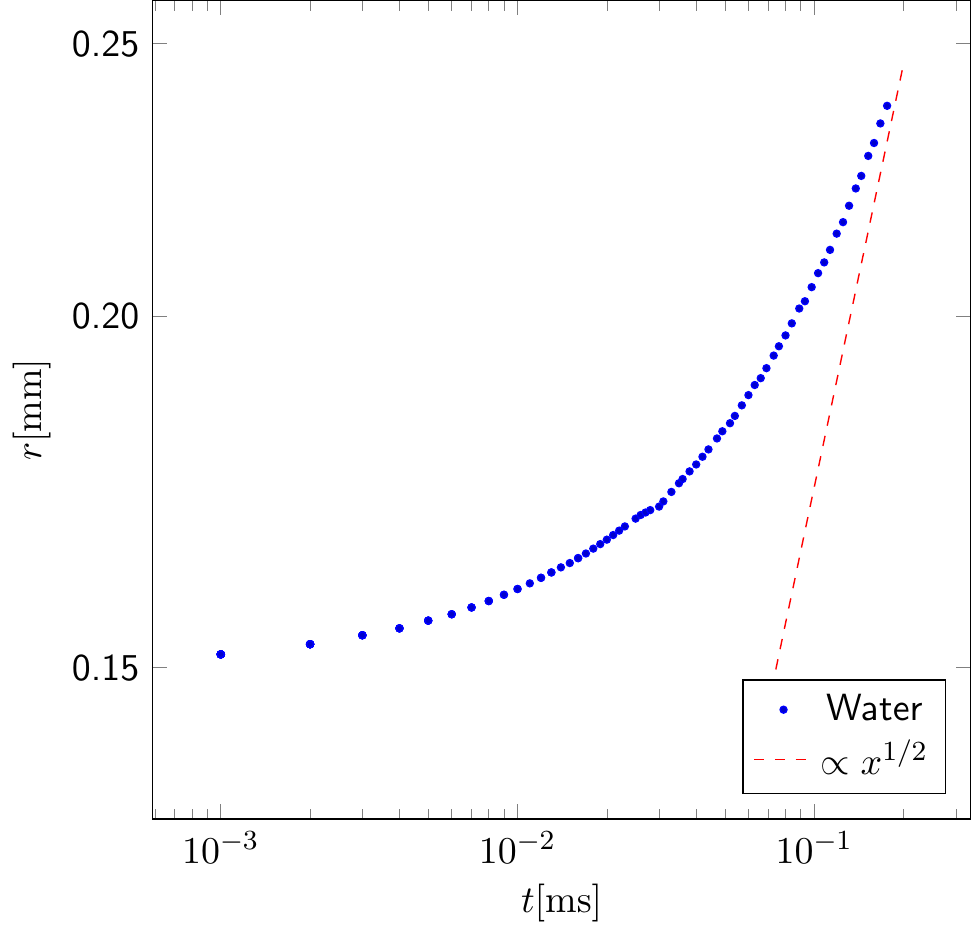}
\caption{\label{figure:diameterTannerRes} Radius of a droplet as a function of time at high resolution.}
\end{figure}
Figure \ref{figure:contourTannerRes} shows the interface ($\alpha = 0.5$) of the
droplet for various times. Figure \ref{figure:diameterTannerRes} shows the
corresponding radial position of the contact line as a function of time. As was
observed in Figure \ref{figure:diameterTanner}, this smaller droplet also shows
inertia dominated spreading.

\begin{figure}
\includegraphics[width=0.6\textwidth]{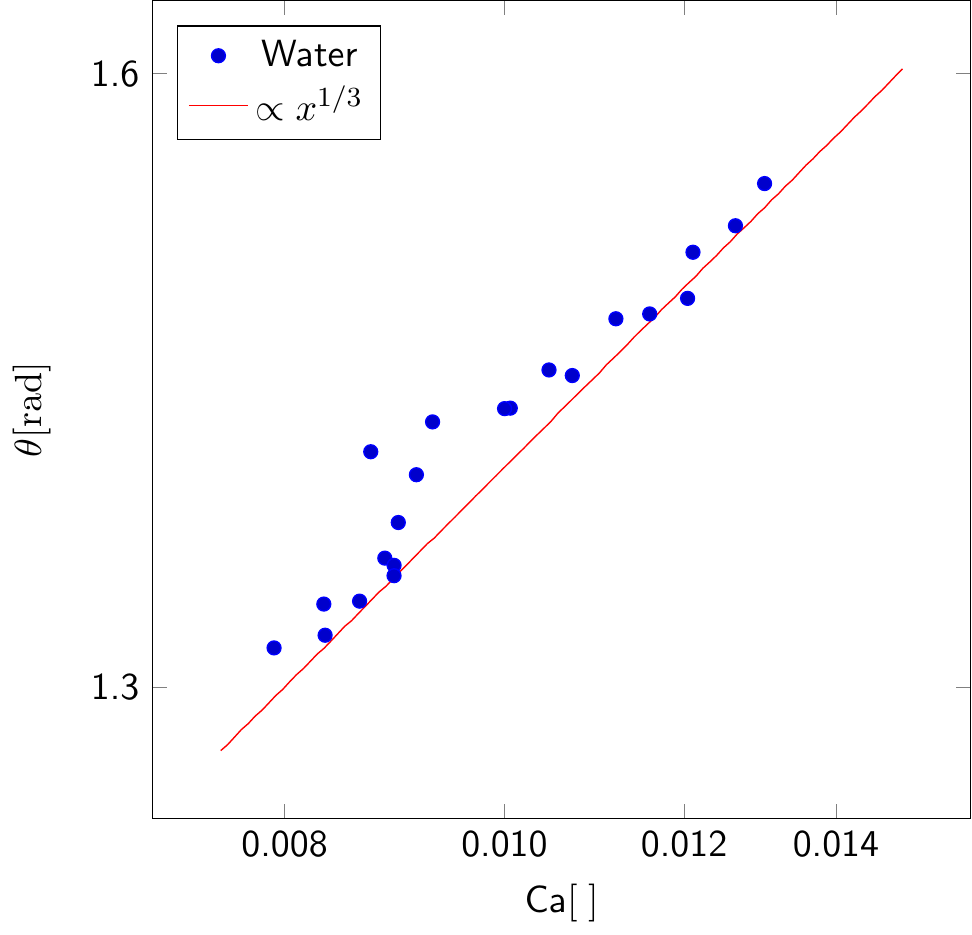}
\caption{\label{figure:angleTannerRes} Angle of the inflection point (i.e. the
apparent contact angle) as a function of the capillary number at high resolution.}
\end{figure}
Because of the larger resolution in these simulations the apparent contact angle
can accurately be determined. Figure \ref{figure:angleTannerRes} shows the
apparent contact angle as a function of the capillary number. It can be
appreciated that the simulation recovers the Voinov-Tanner-Cox
\citep{voinov1976,tanner1979,cox1986} law: ($\theta \propto \txt{Ca}^{1/3}$).

\begin{figure}
\includegraphics[width=0.6\textwidth]{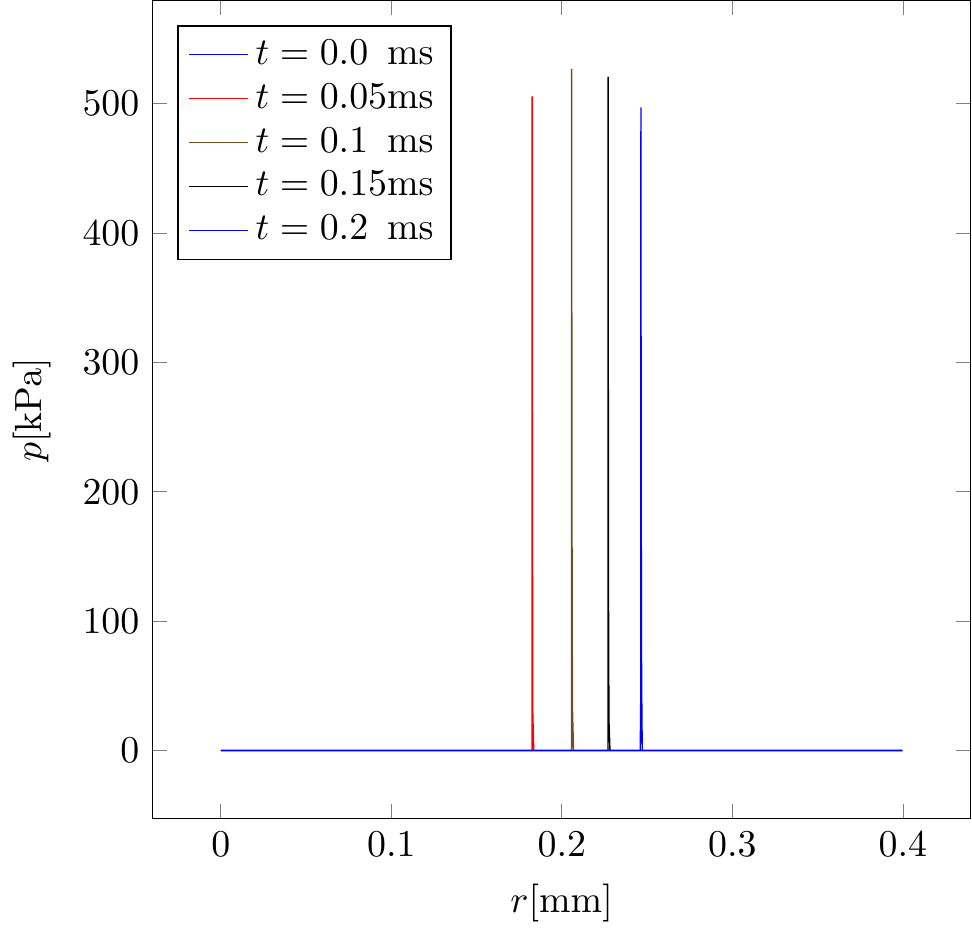}
\caption{\label{figure:pressureTannerRes} Pressure at the wall for various times at high resolution.}
\end{figure}
\begin{figure}
\includegraphics[width=0.6\textwidth]{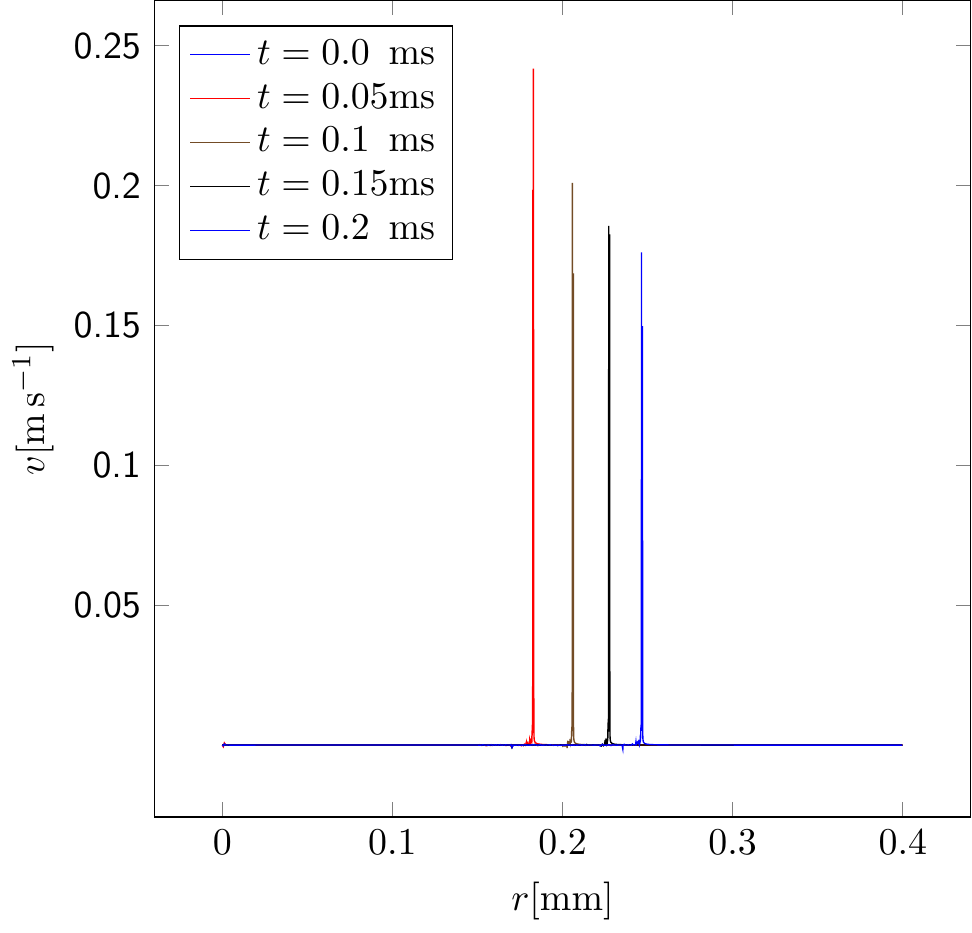}
\caption{\label{figure:velocityTannerRes} Velocity at the wall for various times at high resolution.}
\end{figure}
Figures \ref{figure:pressureTannerRes} and \ref{figure:velocityTannerRes} show
the pressure peak and slip velocity at the wall at various times. Because of the
high resolution at the wall both are very sharp peaks confined to the interface.
The fluctuations of the pressure and velocity that was observed in figures
\ref{figure:pressureTanner} and \ref{figure:velocityTanner} is no longer there,
suggesting the simulations have fully converged.

\section{Conclusions \& Discussion}

An implementation of the Generalized Navier boundary condition for the Volume Of
Fluid method is presented in this work. In analogy with the Brackbill surface
tension model, a body force representation is developed for the contact line
tension, while the Navier slip condition is applied on the wall. A validation of
the code is presented for both a static case of a droplet maintaining its
equilibrium shape, and a dynamic case of a spreading droplet.

It is shown how, on a completely smooth solid surface, in a system without
gravity, the shape of the simulated droplets matches with their theoretically
predicted shape for various Young's angles. In addition, it is shown how the
pressure peak and corresponding velocity peak at the interface converge with
increasing resolution. 

For the dynamic case it is found that the spreading of the droplet scales as $r
\propto t^{1/2}$. This suggests that spreading is limited by inertia. Also the
Voinov-Tanner-Cox law is observed ($\theta \propto \txt{Ca}^{1/3}$). This
behavior is observed without using any fitting parameters.

An interesting future application of this model is flow over pattered surfaces
\citep{dupuis2004,wang2008}. Another topic of interest is the possibility to
investigate both static and dynamic contact angle pinning. One can apply any
pattern of equilibrium contact angles on a surface, and study how different
patterns pin the contact line, either in a static or in a dynamic system.

\begin{acknowledgments}
Many thanks to Sidney Nagel and the Nagel group at the Department of Physics of
the University of Chicago for stimulating discussions and feedback.
\end{acknowledgments}


\end{document}